# DNA-assembled Multilayer Sliding Nanosystems

*Pengfei Zhan[1], Steffen Both[2], Thomas Weiss[2], and Na Liu[1,3,*]*

[1]Max Planck Institute for Intelligent Systems, Heisenbergstrasse 3, D-70569 Stuttgart, Germany

[2]4th Physics Institute and Stuttgart Research Center of Photonic Engineering, University of Stuttgart, 70569 Stuttgart, Germany

[3]Kirchhoff Institute for Physics, Heidelberg University, Im Neuenheimer Feld 227, D-69120 Heidelberg, Germany

[*] Corresponding author. e-mail: na.liu@kip.uni-heidelberg.de

Phone: 0049 711 6891838

**ABSTRACT**

DNA nanotechnology allows for the realization of complex nanoarchitectures, in which the spatial arrangements of different constituents and most functions can be enabled by DNA. When optically active components are integrated in such systems, the resulting nanoarchitectures not only provide great insights into self-assembly of nanoscale elements in a systematic way, but also impart tailored optical functionality to DNA origami. In this Letter, we demonstrate DNA-assembled multilayer nanosystems, which can carry out coordinated and reversible sliding motion powered by DNA fuels. Gold nanoparticles crosslink DNA origami filaments to define the configurations of the multilayer nanoarchitectures as well as to mediate relative sliding between the neighboring origami filaments. Meanwhile, the gold nanoparticles serve as optical probes to dynamically interact with the fluorophores tethered on the filaments, rendering *in situ* detection of the stepwise sliding processes possible. This work seeds the basis to implement DNA-assembled complex optical nanoarchitectures with programmability and addressability, advancing the field with new momentum.

Nature is extremely efficient in creation of biological machines. A class of these biological machines is the motor proteins in living cells[1], which can directly convert chemical energy into mechanical work[2]. Probably, the most studied motor proteins are the members of the kinesin family, which play crucial roles for a broad range of physiological functions[3]. For instance, kinesin-1 is a walking motor protein, which moves in discrete steps along microtubules for intracellular transport[4]. Kinesin-5 is a homotetramer with pairs of motor heads, which crosslink antiparallel microtubules to separate duplicated poles during spindle formation[5]. This results in the relative motion of microtubules, a behavior called microtubule sliding. The key feature of kinesin-5 is its tetramer structure, which allows it to simultaneously move on two microtubules. This feature also distinguishes kinesin-5 from other kinesins, for instance, kinesin-1, which has a dimer structure and binds to one microtubule. In other words, for kinesin-5, the microtubule is both the cargo and the track.

Such natural wonders offer resourceful inspirations and blueprints for constructing DNA-assembled artificial nanosystems[6-12], which mimic the functionality of biological machines[13-17]. Exciting progress has been witnessed, for instance, in the realizations of a DNA walker that programmably collects nanoparticle cargos along an origami assembly line[18], molecular robots guided by prescriptive landscapes[19-21], among others[22-28]. Very recently, we have realized an artificial nanoscopic analog of kinesin-5, in which gold nanoparticles (AuNPs) can mediate sliding of two antiparallel DNA origami filaments powered by DNA fuels[29]. In this Letter, we demonstrate DNA-assembled multilayer sliding nanosystems. Two AuNPs are assembled in between the upper and lower levels of a three-layer nanoarchitecture, respectively. As shown in Figure 1, the two

AuNPs share one common filament in the middle to achieve coordinated motion, so that the DNA filaments in such a system serve as both tracks and cargos for the AuNPs. The stepwise and reversible sliding process is monitored using fluorescence spectroscopy in real time. This work provides great insights into self-assembly of complex dynamic nanoarchitectures with controlled motion on the nanoscale.

Figure 1 illustrates two DNA-assembled multilayer sliding nanosystems labelled as I and II, respectively. In each system, three DNA origami filaments (A, B, C) are grouped together by two AuNPs placed in different levels. In system I, filaments A and C slide in opposite directions with respect to filament B, whereas in system II, filaments A and C slide codirectionally against filament B. Two fluorophores are positioned on filaments A and C, respectively, for *in situ* optically monitoring the multilayer sliding processes, taking advantage of the sensitive distance-dependent interactions between the fluorophores and the AuNPs[30].

Figure 2a shows the schematic and working principle of sliding system I. Three 50 nm-long DNA origami filaments, A (15-helix, light gray), B (14-helix, gray), and C (15-helix, dark gray) are assembled together by folding the M13 scaffolds, staples, and foothold strands. The two ends of each filament are connected through the scaffold strand to ensure the correct relative orientations of the filaments (see the design details in Supporting Information Figure S1 and Table S1). Six rows of footholds evenly separated by 7 nm (coded 1–6) are extended from one lateral side of A (C) and both lateral sides of B. It is noteworthy that the foothold rows are distributed on the neighbouring filaments along their long axes in an antiparallel configuration. There are three binding sites with identical footholds in each row. Two AuNPs (10 nm) crosslink these three filaments in two different levels to define a multilayer system. More specifically, one AuNP is bound in between A and B with four foothold rows, two from each. This emulates the homotetramer structure of kinesin-5, which comprises four motor domains, two on each end. The other AuNP crosslinks B and C in a similar fashion. Two fluorophores (ATTO 550 and ATTO 647N) are tethered on A and C, respectively. To avoid the free rotation of B along its long axis during the assembly process, two side-locks a and b are extended near the ends of the filaments (see Figure 2a). Each DNA lock contains two arms. In side-lock a, one arm tethered on A comprises a 31-basepair (bp) DNA segment (black) with a 12-nucleotide (nt) locking sequence (purple). The other arm positioned on B is composed of its complementary locking sequence (purple) and a toehold segment (gray). Similarly, side-lock b is bridged in between B and C.

To start the sliding process of such a multilayer nanosystem, unlocking strands a′ and b′ are first added to free the structure through toehold-mediated strand displacement reactions[31, 32]. As shown in Figure 2a, the two AuNPs are bound to foothold rows 3 and 4 in both the upper and lower levels. Upon addition of blocking strands 4′ and removal strands $\bar{2}$, through toehold-mediated strand displacement reactions, the two AuNPs are subsequently bound to foothold rows 2 and 3, simultaneously executing one sliding step (see the details in Supporting Information Figure S2). This gives rise to the opposite sliding of A and C relative to B. Figure 2b illustrates the five distinct sliding states of system I. The schematic presented in Figure 2a corresponds to the transition from state iii to state ii in Figure 2b. In this case, the sliding displacement between A and B is -14 nm, which is twice the step size of each AuNP. The sliding displacement between A and C is therefore -28 nm. The sliding process is monitored using fluorescence spectroscopy by *in situ* tracking the fluorescence intensities of ATTO 550 and ATTO 647N using the dual-wavelength time-scan function of a fluorescence spectrometer (Jasco-FP8500) at two emission wavelengths of 578 nm and 663 nm with excitation wavelengths of 550 nm and 647 nm, respectively. The distances of the two fluorophores relative to their respective adjacent AuNPs along the radial directions are given for each state in the upper right (blue) and lower right (red) corners in Figure 2b. As shown in Figure 2c, when transiting from state iii to state ii, the fluorescence intensities of ATTO 550 and ATTO 647N increase (blue) and decrease (red), respectively, as a result of the relative distance enlargement and reduction to their adjacent AuNPs. By sliding one step further along the same direction from state ii to state i, the sliding displacement between A and C becomes as large as -56 nm. Subsequently, with addition of the corresponding DNA fuels, system I can slide to the opposite direction by reversibly passing states ii and iii and completes a full route i-ii-iii-iv-v. As shown in Figure 2c, the relative distance changes between the fluorophores and the AuNPs during the entire sliding process can be optically manifested by the fluorescence intensity changes very well, readily transforming nanoscale motion into optical information.

To validate the experimental observations, theoretical calculations have been carried out, considering the interactions between the fluorophores and the AuNPs. In the weak excitation regime[33], the fluorescence rate $\gamma_{fl}$ of a fluorophore molecule is given as the product of its quantum yield $q$ and its excitation rate $\gamma_{exc}$. Subsequently, the change of the fluorescence rate under the influence of the AuNPs can be written as

$$\frac{\gamma_{fl}}{\gamma_{fl,0}} = \frac{q}{q_0}\frac{\gamma_{exc}}{\gamma_{exc,0}}, \tag{1}$$

in which no subscripts are used to indicate the quantities in the presence of the AuNPs, and the subscripts '0' denote the corresponding quantities in free space. The ratio $\gamma_{exc}/\gamma_{exc,0}$ represents the enhancement of the excitation rate. It is deduced from the finite-element simulations of the near fields generated by a plane wave impinging onto the AuNPs at the wavelengths of 550 nm and 647 nm, respectively. The random orientations of the sliding systems in the solution is taken into account by averaging $\gamma_{exc}/\gamma_{exc,0}$ over all possible incidence directions and polarizations. The quantum yield $q$ in equation (1) can be expressed as

$$q = \frac{\gamma_r/\gamma_{r,0}}{\gamma_r/\gamma_{r,0} + \gamma_{abs}/\gamma_{r,0} + (1-q_0)/q_0}, \qquad (2)$$

in which $\gamma_r$ represents the radiative decay rate in the presence of the AuNPs, $\gamma_{abs}$ is the rate of energy absorption in the AuNPs, and $\gamma_{r,0}$ denotes the radiative decay rate in free space. The factors $\gamma_r/\gamma_{r,0}$ and $\gamma_{abs}/\gamma_{r,0}$ are obtained from the finite-element simulations of an emitting electric dipole placed next to the AuNPs. Special care is devoted to the fact that the fluorophores do not emit at one single wavelength, but over a broad range of wavelengths. This is done by averaging $\gamma_r/\gamma_{r,0}$ and $\gamma_{abs}/\gamma_{r,0}$ over the emission spectra of the fluorophore molecules[34]. The rotational freedom of the fluorophores is taken into account by averaging $\gamma_r/\gamma_{r,0}$ and $\gamma_{abs}/\gamma_{r,0}$ over all possible dipole orientations[35]. As confirmed by our simulations, for the AuNPs as small as 10 nm, the dominating effect is absorption, resulting in quenching of the fluorescence when the molecules approach the metal surface. By comparing the simulations involving both AuNPs and the simulations involving only the AuNP that is closer to the fluorophore (see the structural details in Figure S3), we find that there is no noticeable difference between these two cases for all the calculated rates. Therefore, the influence of the farther AuNP is negligible. The experimental and simulated data show an overall good agreement (see Figures 2c and 2d). The discrepancy is mainly due to the structural imperfections of the sample. More specifically, in the simulation the AuNPs and the DNA origami filaments are modeled with the designed dimensions and perfect shapes. Also, it is assumed that the relative distance changes between the AuNPs and the fluorophores are the same for all the structures, when reaching different stations. In the experiment, however, these parameters cannot be ideal.

In order to successfully characterize the structural properties using transmission electron microscopy (TEM), the DNA-assembled multilayer structures are locked at both ends using eight side-locks (see Figure 3a) to enhance the structural rigidity. This is due to the fact that the free multilayer structures would easily deform after being dried on the TEM grid (see Supporting Information Figure S4 for the DNA origami filament structures without AuNPs and Figure S5 without the side-locks). Figure 3b shows an overview image of the multilayer nanostructures, in which the origami filaments and AuNPs are clearly visible in each structure (see also Supporting Information Figure S6 for additional TEM images). The average TEM image is presented as inset in the same figure. Supporting Information Figures S7 and S8 present the structures of system I before and after sliding, respectively. The displacement between the two AuNPs in the individual structures reveals that system I has successfully carried out relative sliding.

Next, we investigate system II, in which filaments A and C can slide codirectionally with respect to filament B. Figure 4a shows the schematic and working principle of sliding system II. The design of the structure is similar to that in Figure 2a but with two major modifications. First, the foothold rows are distributed in parallel on both of the lateral sides of filaments B. Second, as the foothold row distributions are identical on both of the lateral sides of B, side-locks are not necessary in this case to enforce the correct relative orientations of the filaments. As shown in Figure 4a, the two AuNPs are bound to foothold rows 1 and 2 in the upper and lower levels. Upon addition of blocking strands 1′ and removal strands $\bar{3}$, the two AuNPs both move and are subsequently bound to foothold rows 2 and 3 through toehold-mediated strand displacement reactions. This introduces simultaneous codirectional sliding of A and C against B. The five distinct states of system II are illustrated in Figure 4b. The schematic presented in Figure 4a corresponds to the transition from state i to state ii in Figure 4b. The displacements of A and B with respect to C are shortened from -28 nm to -14 nm. After another sliding step from state ii to state iii, the displacements are decreased to 0 nm in both cases. The sliding process proceeds upon addition of the corresponding DNA fuels. The distances of the two fluorophores relative to their respective adjacent AuNPs along the radial directions are given for each state in the upper right (blue) and lower right (red) corners in Figure 4b. The *in-situ* fluorescence intensities of ATTO 550 and ATTO 647N tracked during the sliding process are presented in Figure 4c by blue and red lines, respectively. The simulated result in Figure 4d agrees well with the experimental observation in Figure 4c.

In conclusion, we have demonstrated DNA-assembled multilayer nanosystems, which can carry out coordinated and reversible sliding motion powered by DNA fuels. The different DNA origami filaments in such multilayer systems can exhibit controlled movements on the nanoscale relative to one another mediated by the AuNPs assembled in between. The motion of the individual filaments have been optically monitored using fluorescence spectroscopy in real time by appropriately introducing distance-dependent interactions between the AuNPs and the fluorophores positioned on the filaments. Our system provides an interesting platform to investigate the mechanic properties of DNA-assembled nanostructures in motion. For instance, studies on the forces exerted during multilayer sliding in dependence on the number of the AuNPs between the origami filaments as well as in the presence of the DNA side-locks will be instructive for understanding the behavior of nanomechanical systems under thermal fluctuations. Also, our work will pave an avenue towards DNA-assembled advanced nanoarchitectures with tailored optical functionality and dynamic complexity.

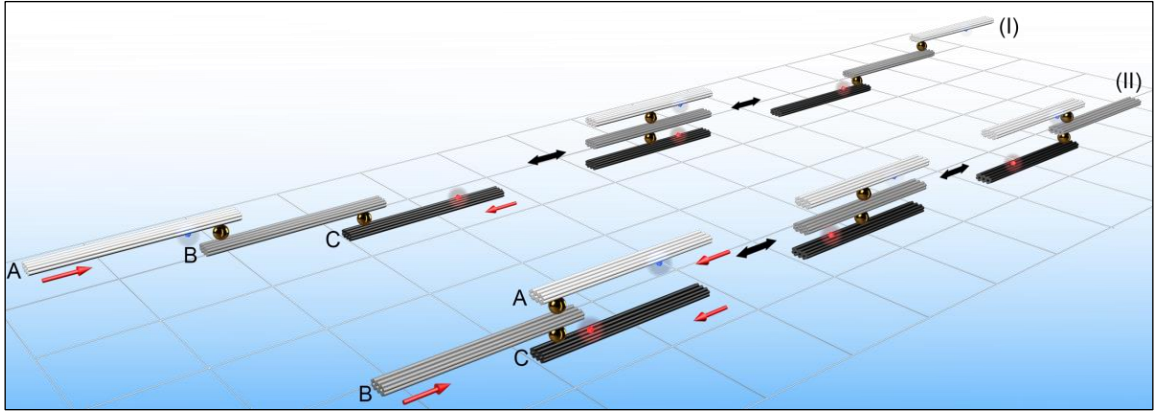

**Figure 1.** Schematics of the DNA-assembled multilayer sliding nanosystems I and II. In each system, three DNA origami filaments (A, B, C) are grouped together by two AuNPs placed in different levels. In system I, filaments A and C slide in opposite directions with respect to filament B, whereas in system II filaments A and C slide codirectionally against filament B. Two fluorophores are positioned on filaments A and C, respectively.

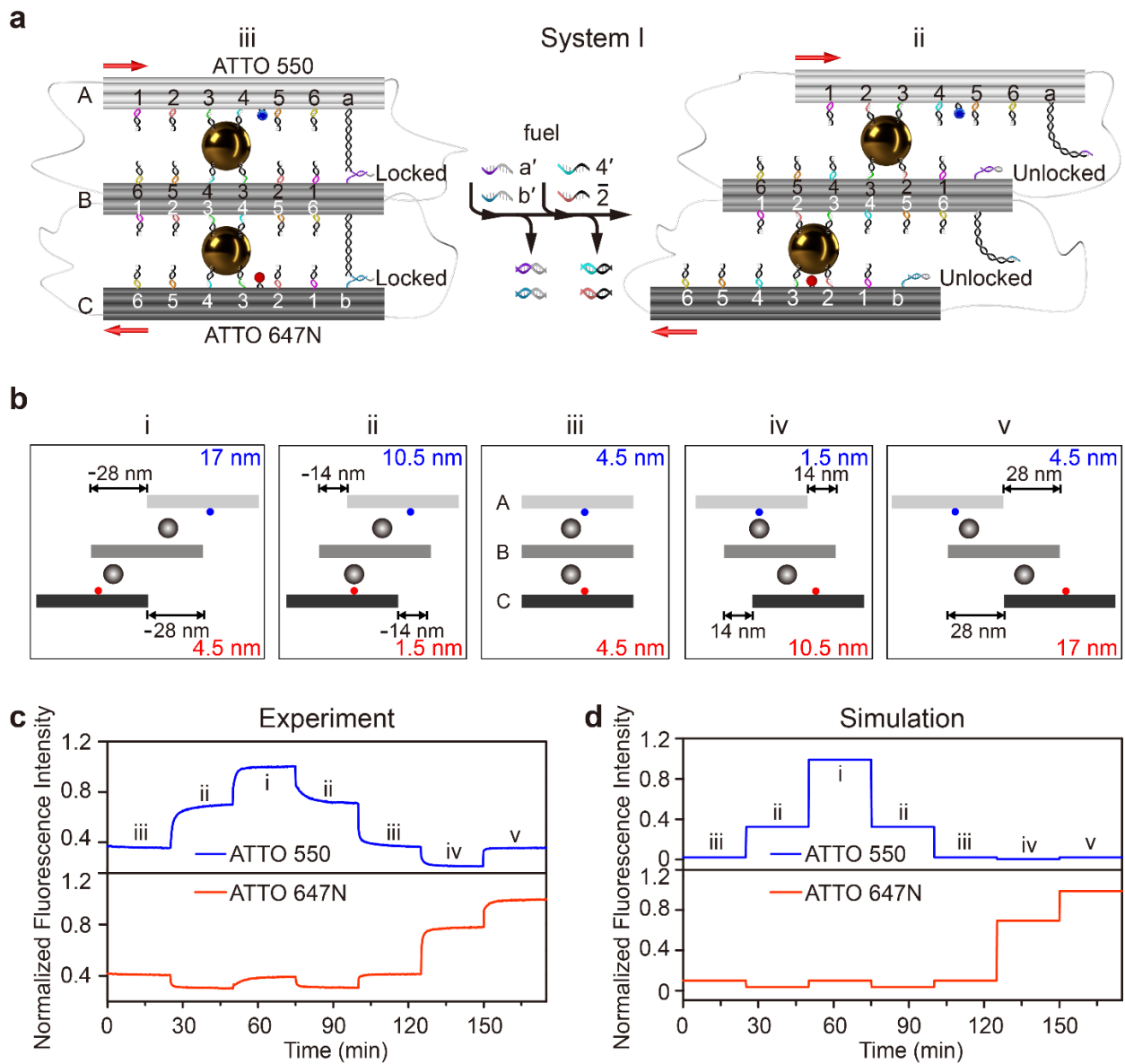

**Figure 2.** Multilayer sliding nanosystem I. (a) Working principle of system I. Six rows of footholds evenly separated by 7 nm (coded 1–6) are extended from one lateral side of A (C) and both lateral sides of B. The foothold rows are distributed on the neighboring filaments along their long axes in an antiparallel manner. There are three binding sites with identical footholds in each row. Two AuNPs (10 nm) crosslink these three filaments in between to define a multilayer system. Two fluorophores (ATTO 550 and ATTO 647N) are tethered on A and C, respectively. To avoid the free rotation of B along its long axis during the assembly process, two side-locks a and b are extended near the ends of the filaments. Upon addition of blocking strands 4′ and removal strands $\bar{2}$, through toehold-mediated strand displacement reactions, the two AuNPs are subsequently bound to foothold rows 2 and 3, simultaneously executing one sliding step. (b) Representative route for the sliding process, comprising five distinct states (i–v). The positions of the two fluorophores and their relative distances to the AuNP surface along the respective radial directions are given for each state. Experimental measurements (c) and theoretical calculations (d) of the fluorescence intensities of ATTO 550 (blue) and ATTO 647N (red) during the siding process from iii to i and then a full route i-ii-iii-iv-v.

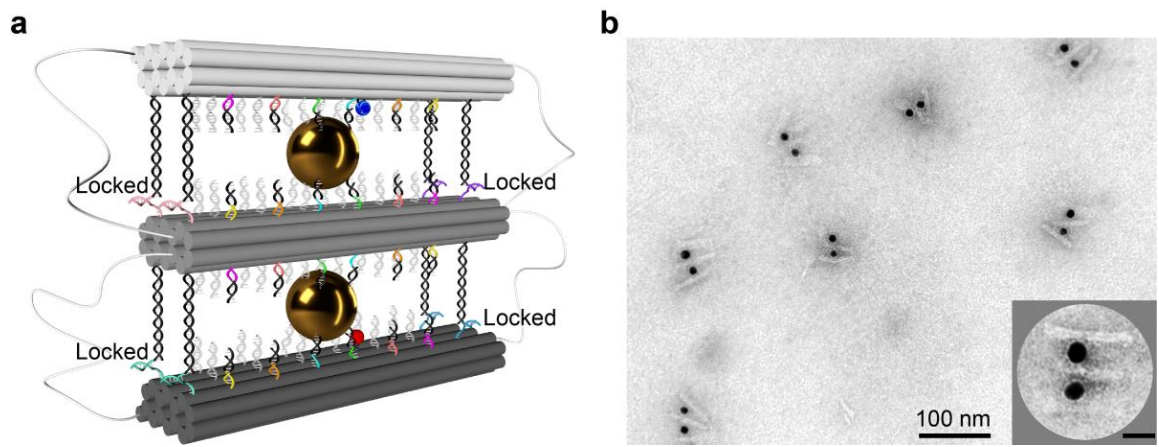

**Figure 3.** (a) Schematic of the multilayer sliding nanosystem that is locked at both ends by eight side-locks to enhance the structural rigidity for TEM structural characterizations. (b) TEM image of the assembled AuNP-origami multilayer sliding structures. Inset: averaged TEM image. Scale bar, 20 nm.

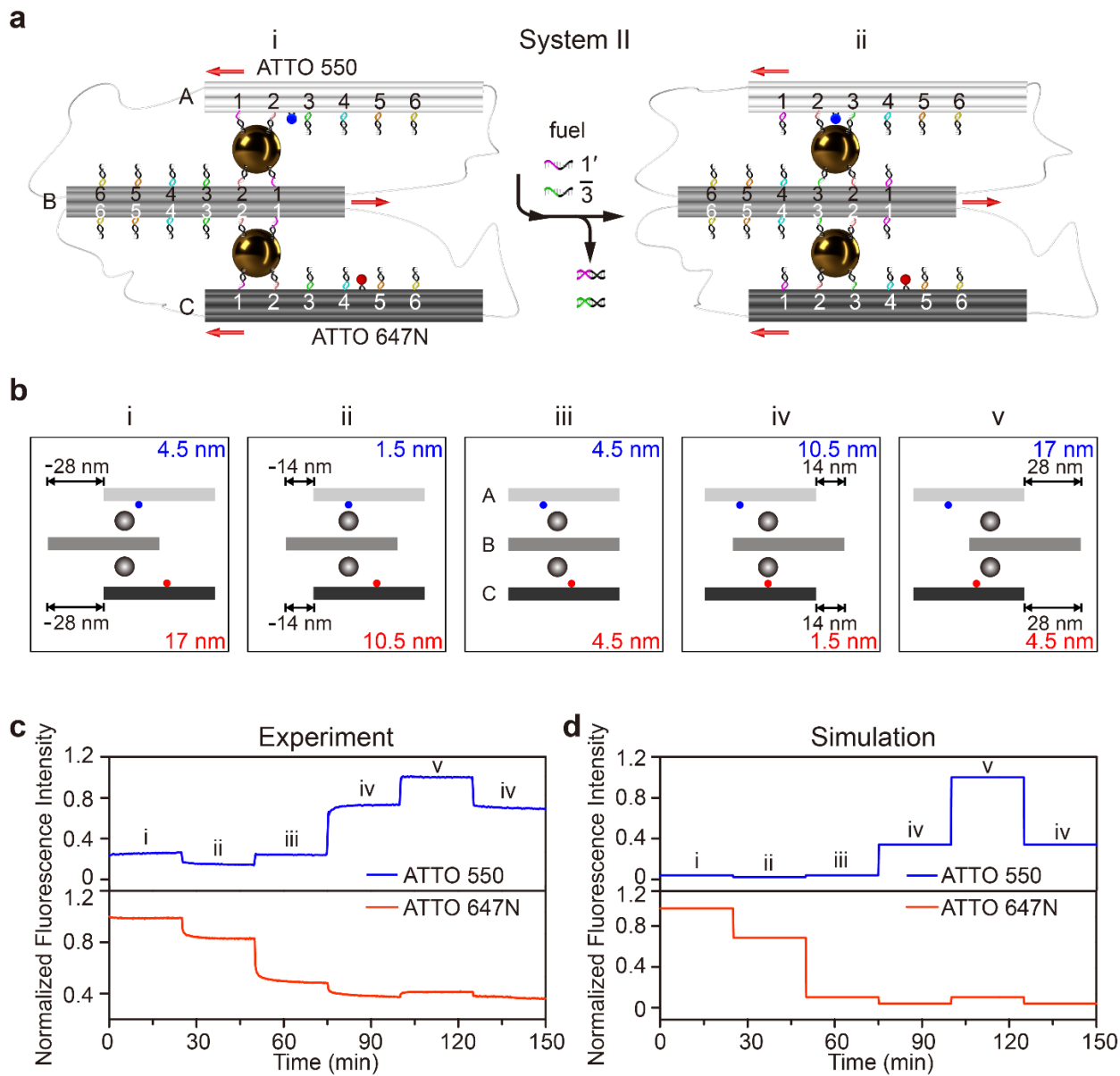

**Figure 4.** Multilayer sliding nanosystem II. (a) Working principle of system II. Upon addition of blocking strands 1′ and removal strands $\bar{3}$, the two AuNPs both move and are subsequently bound to foothold rows 2,3 through toehold-mediated strand displacement reactions. (b) Representative route for the sliding process, comprising five distinct states (i–v). The positions of the two fluorophores and their relative distances to the AuNP surface along the respective radial directions are given for each state. Experimental measurements (c) and theoretical calculations (d) of the fluorescence intensities of ATTO 550 (blue) and ATTO 647N (red) during a sliding route i-ii-iii-iv-v-iv.

**Corresponding Author**

* Corresponding author. e-mail: na.liu@kip.uni-heidelberg.de

**Author Contributions**

N.L. and P.F.Z. conceived the project. P.F.Z. performed all the experiments. S.B. and T.W. carried out the theoretical calculations. P.F.Z., S.B., and N.L. wrote the manuscript. All authors discussed the results and commented on the manuscript.

The authors declare no competing financial interest.

**ACKNOWLEDGMENT**

This project was supported by the Volkswagen foundation and the European Research Council (ERC Dynamic Nano) grant. We thank M. Urban for useful discussions and M. Kelsch for assistance with TEM. TEM images were collected at the Stuttgart Center for Electron Microscopy.

**ASSOCIATED CONTENT**

The Supporting Information is available free of charge on the ACS Publications website at DOI:

Materials and Methods; Figures S1−S8; Tables S1−S3; Supplementary References 1−4

"For TOC only"

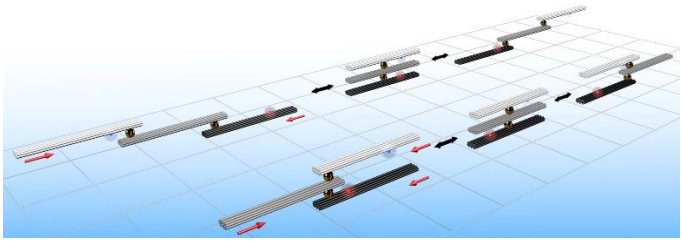